\documentclass[twocolumn,amsmath]{revtex4-2}
\usepackage{amsfonts}
\usepackage{amsmath}
\usepackage{amssymb}
\usepackage{bbm}
\usepackage{bbold} % identity
\usepackage{bm} % bold math
\usepackage{braket}
\usepackage{color}
\usepackage{graphicx} % Include figure files
\usepackage{dcolumn} % Align table columns on decimal point
\usepackage{epsfig}
\usepackage[displaymath,mathlines]{lineno}
\usepackage{natbib}
\usepackage{placeins}
\usepackage{textcomp}
\usepackage{url}
\setcitestyle{super}
\begin{document}
%
%%%%%%%%%%%%%%%%%%%%%%%%%%
% AUTHORS %%%%%%%%%%
%
\author{Marcos Rubín-Osanz$^{1}$, 
Laura Bersani$^{1}$, 
Simone Chicco$^{1}$, 
Giuseppe Allodi$^{1}$, 
Roberto De Renzi$^{1}$,
Athanasios Mavromagoulos$^{2}$,
Michael D. Roy$^{2}$,
Stergios Piligkos$^{2,*}$, 
Elena Garlatti$^{1,3,*}$, 
Stefano Carretta$^{1,3}$}
\email[]{stefano.carretta@unipr.it\\elena.garlatti@unipr.it\\piligkos@chem.ku.dk}
\affiliation{$^1$Dipartimento di Scienze Matematiche, Fisiche e Informatiche, Universit\`{a} di Parma and UdR Parma, INSTM, I-43124 Parma, Italy.\\$^2$Department of Chemistry, University of Copenhagen, DK-2100 Copenhagen, Denmark.\\$^3$INFN, Sezione di Milano-Bicocca, gruppo collegato di Parma, I-43124 Parma, Italy.}
%
%%%%%%%%%%%%%%%%%%%%%%%%%%
% ABSTRACT %%%%%%%%%%%%%%%
%
\begin{abstract}
\noindent
Molecular spin qudits based on lanthanide complexes offer a promising platform for quantum technologies, combining chemical tunability with multi-level encoding.
However, experimental demonstrations of their envisaged capabilities remain scarce, posing the difficulty of achieving precise control over coherences between qudit states in long pulse sequences.
Here, we implement in $^{173}$Yb(trensal) qudit the Quantum Fourier Transform (QFT), a core component of numerous quantum algorithms, storing quantum information in the phases of coherences.
QFT provides an ideal benchmark for coherence manipulation and an unprecedented challenge for molecular spin qudits.
We address this challenge by embedding a full-refocusing protocol for spin qudits in our algorithm, mitigating inhomogeneous broadening and enabling a high-fidelity recovery of the state.
Complete state tomography demostrates the performance of the algorithm, while simulations provide insight into the physical mechanisms behind inhomogeneous broadening.
This work shows the feasibility of quantum logic on molecular spin qudits and highlights their potential.
\end{abstract}
%
%%%%%%%%%%%%%%%%%%%%%%%%%%
% TITLE %%%%%%%%%%%%%%%%%%
\title{Implementation of the Quantum Fourier Transform on a molecular qudit with full refocusing and state tomography}
\maketitle %\maketitle must follow title, authors, abstract, \pacs, and \keywords
%
%%%%%%%%%%%%%%%%%%%%%%%%%%
% INTRODUCTION %%%%%%%%%%%
\section*{Introduction}
%
% Quantum algorithms and qudits
\noindent
The experimental realization of quantum algorithms provides a crucial benchmark for assessing the potential of emerging quantum platforms. 
Most demonstrations have focused on qubit-based architectures such as superconducting circuits \cite{Kjaergaard2020_superconducting, Acharya2023_superconducting, Kim2023_superconducting}, trapped ions \cite{Ballance2016_trapped, Moses2023_trapped, Ringbauer2025_trapped}, and semiconductor quantum dots \cite{Huang2019_silicon, Weinstein2023_silicon, Tanttu2024_silicon}, prompting the second quantum revolution. 
Multi-level quantum systems exploited as qudits---quantum digits with dimension $d > 2$---offer an alternative that has recently gained increasing attention \cite{Blok2021_qutrit, Chi2022_programmable, Ringbauer2022_trapped, Flament2025_coldatom} due to its ability to encode higher-dimensional Hilbert spaces in a single physical system, reducing circuit depth and enabling more efficient quantum error-correction protocols.
Indeed, incorporating quantum error correction and fault-tolerant logic into qudit-based architectures is particularly promising to unlock the full potential of quantum computing in the Noisy Intermediate-Scale Quantum (NISQ) era, where coherence times and gate fidelities remain limited \cite{Shor1996_fault, Knill2005_fault, Devitt2013_qec, Campbell2014_faultqudit, Theral2015_qec, Preskill2018_nisq, Chiesa2020_qec, Mezzadri2023_fault}.

% Molecular spin qudits
Among the available physical platforms, molecular spin qudits (MSQs) represent an appealing route to quantum technologies \cite{MorenoPineda2018_molmagnets, GaitaArino2019_molmagnets, Atzori2019_molmagnets, Chiesa2024_molmagnets} thanks to their chemical tunability, allowing them to have a high number of accessible levels, long coherence times, and potential for scalability through the coupling to superconducting resonators in circuit quantum electrodynamics architectures \cite{Rollano2022_highcoop, Chiesa2023_blueprint}.
In particular, hyperfine-split levels in single-ion systems provide a natural multi-level structure that can be coherently controlled using advanced radio-frequency (RF) pulse sequences \cite{Hussain2018_coherentqudit, Chicco2021_votpp, Chicco2024_simulator}.
The hyperfine manifold of even simple single-ion systems like $^{173}$Yb(trensal) provides, in fact, up to $d=12$ electro-nuclear spin states, with sizeable hyperfine gaps, enabling thermal initialization below 15 mK.
The remarkable potential of MSQs has been thoroughly investigated theoretically \cite{Chiesa2024_molmagnets}, but solid experimental validations of some fundamental steps towards a molecule-based quantum computer are still very few \cite{Godfrin2017_mqsgrover, Chicco2024_simulator, Lim2025_qecmolecule}.
In particular, in a recent work, we provided significant experimental evidence of their capabilities by realizing the first proof-of-concept implementation of a quantum simulator with an ensemble of MSQs \cite{Chicco2024_simulator}.
The accurate quantum simulation of the time evolutions of different target Hamiltonians had the advantage of requiring only the measurement of output populations of the qudit, whereas a demonstration of a full algorithmic control demands the verification of both populations and coherences (off-diagonal elements in the density matrix) of the output state of the qudit, the latter being more prone to errors due to various sources of dephasing.

% What do we do new?
In this work, we experimentally implemented the Quantum Fourier Transform (QFT) on the hyperfine manifold of the molecular spin qudit $^{173}$Yb(trensal).
The QFT is a fundamental primitive in quantum algorithms such as phase estimation and Shor’s algorithm\cite{Eckert1996_qft}, and its high-fidelity execution requires a precise control over coherences across the entire density matrix, where quantum information is stored.
Our molecules are magnetically diluted in a single crystal, ensuring coherence times $T_2 >$ 0.1 ms, while typical driving pulse durations are of a few hundreds of ns for the relevant transitions.
However, working with a molecular ensemble means that inhomogeneous broadening becomes a significant source of dephasing.
In fact, typical characteristic times ($T_2^* <$ 1 $\mu$s) for this effect are comparable to the duration of a short 3-pulse sequence, hindering the direct implementation of the QFT.
We addressed this problem with a full-refocusing scheme tailored for MSQs, embedded in the pulse sequence implementing the QFT, that allows the recovery of coherences even after complex quantum operations consisting of more than 20 pulses.
We demonstrated the performance of our refocused QFT algorithm via complete quantum state tomography, achieving high fidelity of the output qudit state.
Additionally, our measurements and simulations provide new insights into the microscopic mechanisms responsible for inhomogeneous broadening in our MSQs.

% Why it is interesting?
This work represents a key step towards practical implementations of quantum logic operations on MSQs, demonstrating algorithmic control and coherence preservation with high-fidelities in complex pulse sequences.
The sequences designed here, including the refocusing scheme, can be generalized to qudits with $d > 3$ and combined with optimization techniques (e.g., pulse shaping to minimize leakage outside of the desired quantum subspace) \cite{Khaneja2005_nmrgrape, Macedonio2025_bell} to enhance the performance of our hardware.
Thus, our results place molecular spin qudit platforms as promising candidates for qudit-based quantum technologies.
%
%%%%%%%%%%%%%%%%%%%%%%%%%%
% RESULTS %%%%%%%%%%%%%%%%
%
\section*{Results}
%
% CHARACTERIZATION
\subsection*{Characterization and calibration of the qudit}
\begin{figure}[t]
	\includegraphics[width=0.5\textwidth]{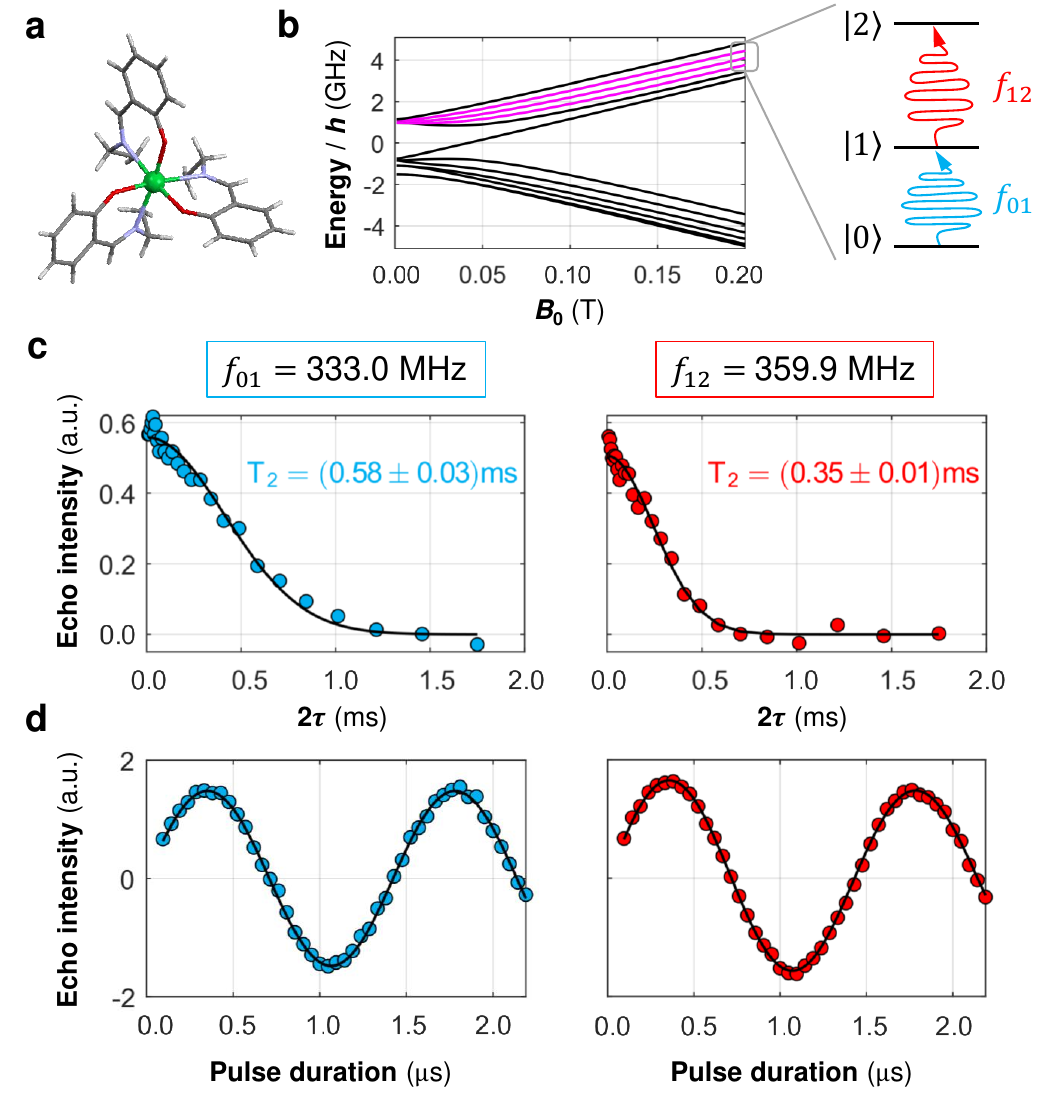}
    \caption{\textbf{Characterisation of the qudit.} 
        \textbf{a} Molecular structure of $^{173}$Yb(trensal).
        \textbf{b} Energy level diagram of $^{173}$Yb(trensal) in a static magnetic field $\textbf{B}_0$ perpendicular to the molecular C$_3$ axis.
        The states of the subspace targeted in this work, highlighted in pink, are labelled $\ket{0}$, $\ket{1}$ and $\ket{2}$ from lowest to highest energy, with transition frequencies $f_{01}$ and $f_{12}$.
        \textbf{c} Phase memory time $T_2$ of both transitions, measured at $B_0=0.2$ T and $T=1.4$ K (scatters) and fit to a Gaussian-shaped decay (solid line).
        \textbf{d} Coherent Rabi manipulation of both transitions under these same experimental conditions.
        Error bars on all data are of the order of scatter dimension; they were obtained by applying to the noise the same analysis we applied for the data (see Methods) and taking the root mean square.
    }
	\label{fig1:characterization}
\end{figure}

%
% The 173Yb(trensal) system
\noindent
Our platform is an isotopically enriched single crystal of $^{173}$Yb(trensal), diluted at 0.05$\%$ in its diamagnetic isostructural analogue Lu(trensal) to reduce intermolecular magnetic interactions, see Methods.
At this dilution, each $^{173}$Yb(trensal) molecule \cite{Pedersen2015_ybtrensal, Pedersen2016_ybtrensal} (Fig. \ref{fig1:characterization}a) behaves as an isolated qudit \cite{Hussain2018_coherentqudit} consisting of an effective electronic spin $S=1/2$ coupled to the nuclear spin $I=5/2$ of the $^{173}$Yb isotope, for a total of 12 energy levels (Fig. \ref{fig1:characterization}b) described by the effective Hamiltonian (in frequency units)
\begin{equation}
    \begin{split}
        H_0 = &\,\,  
        A_\parallel S_z I_z + A_\perp (S_x I_x +S_y I_y) + p I_z^2 \\
        &\,\, + \left( \dfrac{\mu_{\rm B}}{h} \right) 
        \textbf{S} \cdot \textbf{g} \cdot \textbf{B}_0 
        - \left( \dfrac{\mu_{\rm N}}{h} \right) 
        g_I \textbf{I} \cdot \textbf{B}_0 \, .\\ 
    \end{split}
    \label{eq:spin_Hamiltonian}
\end{equation}
In Equation (\ref{eq:spin_Hamiltonian}), $\textbf{S}$ and $\textbf{I}$ represent the electronic and nuclear spin operators, respectively (with components on the molecular axes $\hat{x}, \hat{y}, \hat{z}$) and the first two terms are their strong hyperfine interaction, axial on the $C_3$ axis ($\hat{z}$) of symmetry of the molecule, with hyperfine constants $A_\parallel = -883$ MHz and $A_\perp =-628$ MHz.
The third term describes the nuclear quadrupolar coupling ($p = -66$ MHz), while the last two are the electronic and nuclear Zeeman couplings to an external static magnetic field $\textbf{B}_0$, with the electronic coupling being axial on the $C_3$ axis ($g_x=g_y=2.9$, $g_z=4.3$) and the nuclear one isotropic ($g_I=-0.2592$) \cite{Hussain2018_coherentqudit, Rollano2022_highcoop}.
These parameters generate an energy spectrum with distinct transition frequencies that allow both a selective, single transition manipulation and a multiple transition addressing in a broadband setup.

% FIGURE 1: Characterization

% Characterization of the target subspace, coherent control
For the implementation of the QFT algorithm, we set a magnetic field $B_0 = 0.2$ T along $\hat{x}$, orthogonal to the $C_3$ axis, for which the electronic Zeeman is the dominant interaction in Eq.(\ref{eq:spin_Hamiltonian}).
This field splits the 12 states into two energy manifolds, with each state almost factorized in the electronic and nuclear spin subspaces, allowing a simple labeling in terms of the electronic and nuclear spin components along $\hat{x}$: $\ket{m_S,m_I}$.
We selected the 3-level subspace of the upper energy manifold highlighted in pink in Fig. \ref{fig1:characterization}b, which consisted of states $\ket{0}=\ket{+\frac{1}{2},+\frac{1}{2}}$, $\ket{1}=\ket{+\frac{1}{2},-\frac{1}{2}}$, and $\ket{2}=\ket{+\frac{1}{2},-\frac{3}{2}}$.
The two addressable transitions $\ket{0} \leftrightarrow \ket{1}$ and $\ket{1} \leftrightarrow \ket{2}$, with frequencies $f_{01}= 333.0$ MHz and $f_{12}= 359.9$ MHz, were the only ones within the frequency bandwidth ($320-370$ MHz) of the custom probe designed for the experiment, mitigating leakage to states outside this subspace.
The system was cooled to $T=1.4$ K, where we measured the relaxation ($T_1$)---see Supplementary Information, SI---and coherence ($T_2$) times of both transitions.
A remarkable feature of $^{173}$Yb(trensal) at this dilution and temperature is its long coherence times ($T_2 >$ 0.1 ms), as shown in Fig. \ref{fig1:characterization}c.
Thus, the effect of pure dephasing in our results is small, as $T_2$ is much longer than sequence duration.
The Gaussian shape of the decay of coherences is expected when dipolar interactions are the dominant dephasing mechanism \cite{Ratini2025_decoherencequdits}.

% CONTROL AND TOMOGRAPHY
\subsection*{Coherent control and qutrit state tomography}
\begin{figure}[t]
	\includegraphics[width=0.5\textwidth]{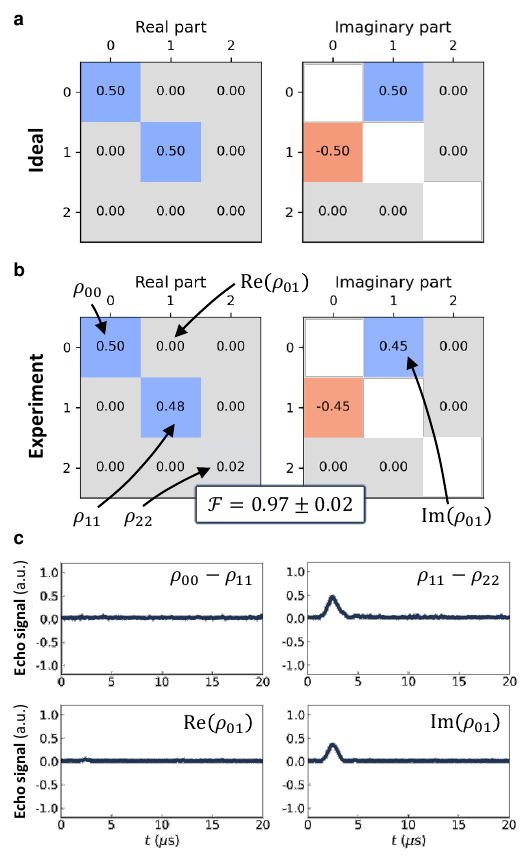}
    \caption{\textbf{Coherent manipulation and tomography of the qutrit state.}
        \textbf{a.} Density matrix $\rho_{\rm ideal} = \ket{\psi}\bra{\psi}$ for the ideal superposition state $\ket{\psi} = \tfrac{1}{\sqrt{2}} (\ket{0} - i\ket{1})$.
        \textbf{b.} Density matrix $\rho_{\rm exp}$ for the final state after the experimental manipulation of the initial state of the qutrit, $\ket{0}$, with a $\theta=\pi/2$ pulse with phase $\phi=0$ and carrier frequency $f_{01}$.
        This pulse ideally generates the superposition state of panel a.
        The elements of $\rho_{\rm exp}$ were measured by performing a complete tomography experiment of the final state.
        \textbf{c.} Measured echo traces yielding the tomography of panel b.
        We estimated the error in $\mathcal{F}$ by applying the same noise analysis of Fig. \ref{fig1:characterization} to these traces.
    }
	\label{fig2new}
\end{figure}

\noindent
Coherent control was achieved via RF pulses, sent with an Arbitrary Waveform Generator (AWG) by Active Technologies and the home-made broadband spectrometer \emph{HyReSpect} \cite{Allodi2005_spectrometer}, as demonstrated by the Rabi curves in Fig. \ref{fig1:characterization}d.
Each pulse ideally produces a planar rotation in one of the two target transitions, defined as
\begin{equation} \label{eq:planar_rotation}
    \begin{split}
    P_{\mu\nu} \left(\theta,\phi\right) = 
    &\cos{\left( \dfrac{\theta}{2} \right)} 
    \left( \ket{\mu}\bra{\mu} 
    + \ket{\nu}\bra{\nu} \right) \\
    & - i \sin{\left( \dfrac{\theta}{2} \right)} 
    \left( e^{i \phi} \ket{\nu}\bra{\mu} 
    + e^{-i \phi} \ket{\mu}\bra{\nu} \right)
    \, , \\
    \end{split}
\end{equation}
\noindent implementing a single qutrit gate involving states $\ket{\mu}$ and $\ket{\nu}$.
The tilt angle $\theta$ of the rotation is controlled by the RF power, duration and shape of the pulse; the phase $\phi$ is that of the pulse carrier signal of frequency $f_{\mu\nu}$.
As an example, the ideal effect of a single pulse with $\theta=\frac{\pi}{2}$, $\phi=0$ and carrier frequency $f_{01}$ on state $\ket{0}$, which generates the superposition state $\ket{\psi} = P_{01} \left(\pi/2,0\right) \ket{0} = \tfrac{1}{\sqrt{2}} \left( \ket{0} - i\ket{1} \right)$, is shown in Fig. \ref{fig2new}a as the density matrix $\rho_{\rm ideal} = \ket{\psi}\bra{\psi}$.
Our experimental implementation of the same gate yielded the density matrix $\rho_{\rm exp}$ in Fig. \ref{fig2new}b, obtained by performing a full tomography experiment on the final superposition state.
This demonstrates our coherent control of the qutrit state beyond Rabi experiments, with a gate fidelity\cite{Nielsen_quantcomp:ch9} $\mathcal{F} = \rm{Tr} \left( \sqrt{ \sqrt{\rho_{\rm ideal}} \rho_{\rm exp} \sqrt{\rho_{\rm ideal}}} \right) = 0.97\pm0.02$.
Figure \ref{fig2new}c shows the echo traces used to reconstruct $\rho_{\rm exp}$. See Methods for more details on the tomography experiments.

% FIGURE 2 new: Tomography example

As the temperature of the system was still too high for a thermal initialization to a pure state, we generated a pseudo-pure state before every state manipulation experiment\cite{Chicco2024_simulator}.
To do so, we sent a $\frac{\pi}{2}$ pulse with $f_{12}$, equipopulating states $\ket{1}$ and $\ket{2}$, and let the coherence between the two states decay for a time $t \gg T_2$.
The resulting density matrix can be then decomposed into a term proportional to the identity matrix, unaffected by further manipulation, and a term proportional to a pure state.
Thus, measurements on a pseudo-pure state are effectively equivalent to those on a pure state up to a normalization factor (see Methods).

% QFT IMPLEMENTATION
\subsection*{Implementation of the QFT}

% FIGURE 3: QFT sequences
\begin{figure*}[t]
	\begin{centering}
		\includegraphics[width=1.0\textwidth]{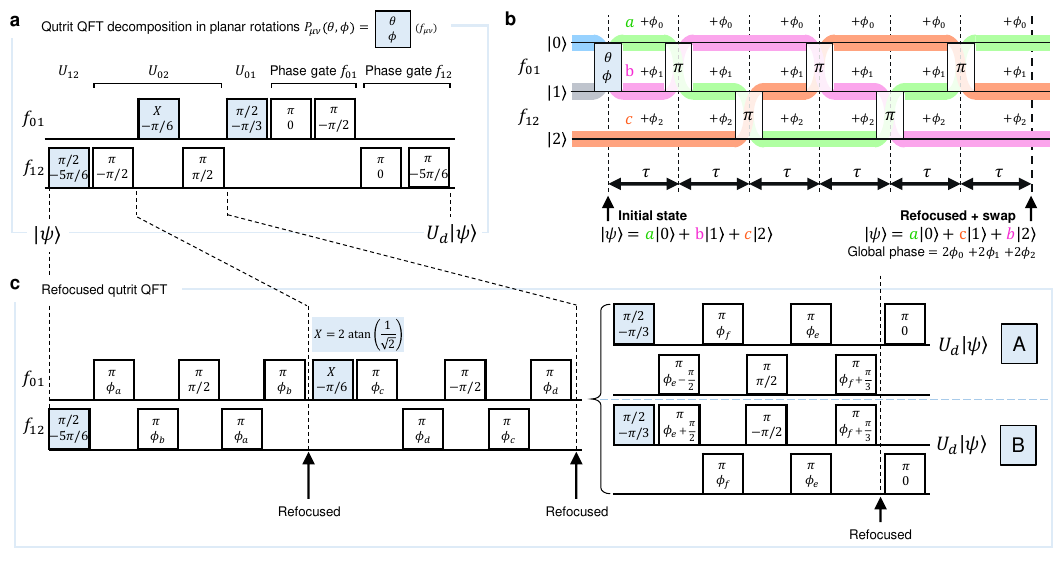}
		\caption{{\bf Pulse sequence for the QFT}. 
            \textbf{a} Decomposition of the 3-level unitary transformation $U_d$ performing the qutrit QFT into 9 planar rotations. 
            \textbf{b} Detail of a qutrit refocusing block.
            After a pulse ($\theta$, $\phi$) of the QFT sequence, a state $\ket{\psi} = a \ket{0} + b \ket{1} + c \ket{2}$ is encoded into the three-level subspace with amplitudes $a$ (green), $b$ (pink), and $c$ (orange). 
            Due to inhomogeneous broadening, the free evolution of each spin in the ensemble is slightly different, and each amplitude of each spin collects a different phase during the time $\tau$ between pulses.
            This phase depends on the basis state ($\ket{0}$, $\ket{1}$, $\ket{2}$) in which the amplitude is encoded ($\phi_0$, $\phi_1$ and $\phi_2$,  respectively).
            A $\pi$ pulse swaps the amplitudes between the basis states.
            By applying a sequence of five $\pi$ pulses, equally spaced in time, amplitudes are moved around all the basis states, thus collecting the same global phase $2 \phi_0 + 2 \phi_1 + 2 \phi_2$.
            This final refocused state includes a swap between two states.
            \textbf{c} Pulse sequence implementing $U_d$ with embedded refocusing blocks (see panel b) for any pulse with $\theta \neq \pi$.
            Any such pulse must be sent to a refocused state after a refocusing block.
            Conversely, $\pi$-pulses can be integrated in them to reduce the number of pulses.
            Only relative phases between some of the $\pi$-pulses are set by the QFT operation, we can freely chose $\phi_a$, $\phi_b$, $\phi_c$, $\phi_d$, $\phi_e$ and $\phi_f$.
            All results here were obtained with $\phi_a = \phi_b = \phi_c = \phi_d = \phi_e = 0$ and $\phi_f = -\pi/6$.
            Detection was performed when the density matrix was refocused, before the last $\pi$ pulse.
            Both sequences A and B implement the QFT, each one allowing for the detection of different density matrix elements (see Methods).
        }
		\label{fig3new}
    \end{centering}
\end{figure*}

% Original QFT sequence: effect of inhomogeneous broadening
\noindent The QFT operation on a qudit with $d$ states is described by the unitary transformation
\begin{equation} \label{eq:qft_matrix}
U_d = \frac{1}{\sqrt{d}}\sum_{\mu,\nu=0}^{d-1}e^{i2\pi\frac{\mu\nu}{d}}\ket{\mu}\bra{\nu}
\, .
\end{equation}
The implementation of $U_d$ for a qutrit ($d = 3$), depicted in Fig. \ref{fig3new}a, consists of a sequence of optimized RF pulses based on its decomposition into planar rotations $P_{\mu\nu} \left(\theta,\phi\right)$ in the two addressable transitions \cite{Chizzini2022_qftrotations}.
However, measuring the result of this sequence in an ensemble of MSQs is greatly affected by inhomogeneities in the sample: coherences are lost in a timescale $T_2^* \ll T_2$ during the computation.
For $^{173}$Yb(trensal), in the dilution and temperature conditions of these experiments, both transitions had $T_2^* \sim 500$ ns, determined from the width of the measured echoes (see SI).
This is comparable to the timescale of our pulses, and therefore significantly shorter than the whole sequence (9 pulses), reducing the fidelity of the implementation.
In this regard, the QFT is a very stringent test for experiments with a molecular ensemble, as it generates all-to-all coherences and encodes quantum information in their relative phases.

% FIGURE 4: QFT results
%
\begin{figure*}[t]
	\begin{centering}
		\includegraphics[width=1.0\textwidth]{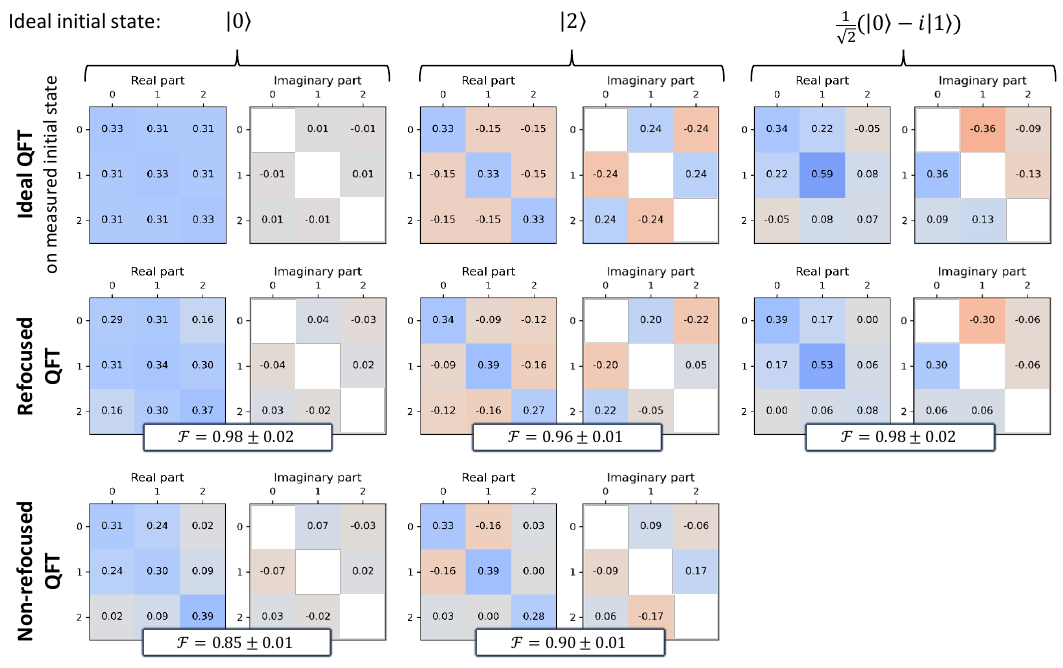}
		\caption{\textbf{Tomography of the QFT acting on the basis states $\ket{0}$ and $\ket{2}$ of the qutrit, and on the superposition state of Fig. \ref{fig2new}.} 
        We compared the density matrices $\rho_{\rm exp}$ obtained from tomography experiments after implementing the QFT sequences (both refocused and non-refocused) with the density matrix $\rho_{\rm ideal}$ obtained by applying $U_d$ to the density matrix of the initial state ($\rho_0$) extracted from tomography experiments (see SI for the tomography of $\rho_0$). For each $\rho_{\rm exp}$ we calculated the fidelity as $\mathcal{F} = \rm{Tr} \left( \sqrt{ \sqrt{\rho_{\rm ideal}} \rho_{\rm exp} \sqrt{\rho_{\rm ideal}}} \right)$, which measures only the performance of pulse sequence implementing the QFT. 
        An error in the order of 0.01 was obtained for all matrix elements from the standard deviation of repeated measurements. The error on the fidelity was obtained by propagating the single element errors.}
        \label{fig4}
	\end{centering}
\end{figure*}

% Including refocusing
The effect of inhomogeneous broadening can be mitigated by implementing a refocusing scheme for qutrits\cite{Vitanov2015_quditrefocusing}. 
An example of a sequence that refocuses the qutrit state is depicted in Fig. \ref{fig3new}b: after applying an arbitrary pulse, which generates a state $\ket{\psi} = a \ket{0} + b \ket{1} + c \ket{2}$, a sequence of five $\pi$-pulses alternating between the two addressable transitions transforms any phases generated by the free evolution of each state of each spin into a global phase, and swaps two states.
A similar strategy has been implemented on qutrits encoded in cold atoms and superconducting processors to refocus a previously generated superposition qutrit state \cite{Yuan2022_refocusing, Iiyama2024_refocusing, Tripathi2025_refocusing}.
However, our relatively short $T_2^*$ requires embedding the refocusing in the manipulation of the qutrit state.
For this, we designed the 19-pulse sequences in Fig. \ref{fig3new}c, which implement the QFT while refocusing the qutrit state every 6 pulses.
Numerical simulation of the full pulse-sequence on a ensemble of $^{173}$Yb(trensal) molecules, each one described by the Hamiltonian of Eq.(\ref{eq:spin_Hamiltonian}) plus a strain in the Hamiltonian parameters (sampled from a Gaussian distribution around their mean values), yielded an almost perfect recovery of the ideal QFT operation in the absence of other sources of errors (see SI for simulated tomography results).
We also found that the relatively long pulses used in the detection sequence act as a filter, suppressing contributions from spins exhibiting the strongest dephasing due to inhomogeneous broadening.
As a result, the measurement predominantly probes spins located near the centre of the distribution, for which the refocusing procedure is most effective.

We experimentally implemented the QFT, both with the standard (Fig. \ref{fig3new}a) and the refocused (Fig. \ref{fig3new}c) pulse sequences, on the three basis states.
We also tested the refocused sequence with two different superpositions as initial states.
Figure \ref{fig4} shows the tomography ($\rho_{\rm exp}$) of the final states after our implementation of the QFT on two of the basis states ($\ket{0}$, $\ket{2}$) and the superposition state of Fig. \ref{fig2new} (see SI for the other two initial states).
In the same figure, we compare these results with the density matrix $\rho_{\rm ideal}$ obtained by applying the ideal QFT operation represented by $U_d$ to the $\rho_{\rm exp}$ obtained from the tomography of the initial states (reported in the SI).
In this case, the quantity $\mathcal{F} = \rm{Tr} \left( \sqrt{ \sqrt{\rho_{\rm ideal}} \rho_{\rm exp} \sqrt{\rho_{\rm ideal}}} \right)$ represents the fidelity of the refocused QFT alone.
Similar results for the initial basis state $\ket{1}$ and superposition $\frac{1}{\sqrt{2}}(\ket{1}-i\ket{2})$) are reported in the SI.
Tomography results after the pristine QFT sequence without refocusing reveal that the populations of the qutrit’s final state are accurately reproduced, whereas the coherences---particularly the two-quanta ones---are substantially attenuated, yielding fidelities $\mathcal{F}\leq 0.9$ (e.g., $\mathcal{F} = 0.85\pm0.01$ for the initial state $\ket{0}$ and $\mathcal{F} = 0.90\pm0.01$ for $\ket{2}$ ).
Conversely, as one can evince from Fig. \ref{fig4}, the implementation of the refocusing scheme allows for almost a full recovery of all coherences across the density matrix and a consequent improvement of the fidelity $\mathcal{F}\geq 0.96$ (e.g., $\mathcal{F} = 0.98\pm0.02$ for $\ket{0}$, $\mathcal{F} = 0.96\pm0.01$ for $\ket{2}$, $\mathcal{F} = 0.98\pm0.02$ for $\frac{1}{\sqrt{2}}(\ket{0}-i\ket{1})$), despite the very long and complex sequence of pulses. 

%%%%%%%%%%%%%%%%%%%%%%%%%%
% DISCUSSION %%%%%%%%%%%%%
%
\section*{Discussion}
\noindent
The results demonstrate that we implemented the Quantum Fourier Transform on a molecular spin qudit with a complex sequence of RF pulses, including a refocusing scheme and a full tomography of the final quantum state.
The high fidelities resulting from our QFT algorithm mark a significant step forward in the field of MSQs.
Indeed, these unprecedented results demonstrate that it is possible to implement complex quantum algorithms on MSQs, while retaining precise control over both populations and coherences.
In particular, the refocusing scheme designed and implemented in this work has proven to be an effective strategy for suppressing the effects of inhomogeneous broadening \emph{on the fly} while implementing the algorithm.
We demonstrated experimentally that this scheme mitigates the source of error from inhomogeneous broadening and this potentially sets the long characteristic time of pure dephasing in $^{173}$Yb(trensal) as the time limit for the experiment.
Our results constitute, to the best of our knowledge, also the first implementation of full refocusing in molecular spin qudits, opening the door to implement other qudit-based algorithms in this platform even in the presence of inhomogeneous broadening.
In addition, the refocusing strategy can be easily adapted to other algorithms and qudit dimensions.
In this regard, we note that the implementation of the refocusing sequence could be optimized by using a system with more connectivity.
For example, refocusing a system with $d$ even and loop connectivity---linear connectivity with an additional connection between the edges---would require sending half the number of pulses than with linear connectivity.
\\
\indent
Simulations of our pulse sequence---both refocused and non-refocused---on our ensemble of $^{173}$Yb(trensal) molecules confirm our experimental results. 
We simulated the action of our pulse sequences with the experimental parameters, also including the detection scheme.
We found that with the non-refocused sequence, coherence is partially lost due to inhomogeneous broadening, as $T_2^*$ for both transitions is much shorter than the sequence duration.
In contrast, the ideal simulation of the refocused sequence in the absence of other sources of error completely recovers all coherences. 
We also employed our simulations to check the effect of introducing a strain in different parameters of the spin Hamiltonian as a source of inhomogeneous broadening.
From the observed $T_2^*$ in the two addressable transitions in this work, we conclude that the dominant mechanism is a strain in the hyperfine couplings (see SI).
\\
\indent
As a last remark, we note that refocusing strategies for qudits are used in other quantum platforms due to an inhomogeneous broadening arising between repetitions of the same experiment in the same quantum system.
In transmons \cite{Tripathi2025_refocusing} this effect is intrinsic to the system, arising from the additional noisy excited state used to encode the qutrit (transmons are optimized to work as qubits).
In contrast, we implemented a refocusing strategy to mitigate the inhomogeneous broadening coming from working with a large ensemble of molecules.
As molecular spins can naturally encode qudits, a more resilient behaviour is expected when going to the single molecule limit, since the dominant source of inhomogeneous broadening in our sample is a statistical distribution of hyperfine couplings.
Nonetheless, refocusing schemes like the one in this work can still represent a crucial tool to mitigate the effects of other sources of inhomogeneous broadening, like the Overhauser field due to the coupling with surrounding nuclear spins in the molecule \cite{Chiesa2021_dephasing}.

%%%%%%%%%%%%%%%%%%%%%%%%%%
% METHODS %%%%%%%%%%%%%%%%
%
\section*{Methods}
%
% SAMPLE
\subsection*{Sample}
\noindent
Three separate batches of single crystals of $^{173}$Yb(trensal) magnetically diluted in Lu(trensal) were prepared\cite{Chicco2024_simulator} with a nominal dilution of 0.05$\%$ $^{173}$Yb@Lu.
Inductively coupled plasma mass spectrometry (ICP-MS) was used to analise the dilution, using an identical procedure for all three batches.
Between 3.85 and 4.03 mg of the $^{173}$Yb@Lu(trensal) product were digested in 1.0 mL trace analysis nitric acid and diluted to 10.0 mL.
A portion of this solution was diluted by half to reach a concentration equivalent to 26.40--27.63 parts-per-billion (ppb) $^{173}$Yb.
Another portion of the initial solution was diluted by 2000 times (serially 1/20 and 1/100) to afford a concentration equivalent to 53.39--55.89 ppb Lu.
ICP-MS results based on a calibration curve, scaled by 0.161/0.99 to reflect the enrichment with $^{173}$Yb from 16.1$\%$ (natural abundance) in the calibrant to 99$\%$ in the sample, found: 63.5591--62.1965 ppb $^{173}$Yb and 40.4394--58.0067 ppb Lu, corresponding to a molar doping percent of 0.0416(6)$\%$-0.0592(8)$\%$ $^{173}$Yb@Lu.
Error was propagated from the dilution ratios, neglecting errors in instrument calibration and isotope ratios.

% PULSE GENERATION
\subsection*{Pulse generation}
\noindent
The experimental setup consisted of two different pulse generators---a 6 GS/s AWG by Active Technologies, model AWG-5062D, and the \textit{HyReSpect} spectrometer---that sent pulses to a probehead (an LC circuit with a wide bandwidth) where the sample was located.
We chose the orientation of the sample in the probe to have the driving RF magnetic field of the pulses applied along the molecular axis $\hat{z}$, perpendicular to the static magnetic field.
The pulses required for the QFT sequence were generated by the AWG, whereas the Hahn-echo excitation was performed by the spectrometer.
The heterodyne-based detection performed by the latter was coherent with its pulse carrier, while the spectrometer and the AWG were mutually incoherent.
RF output power was tuned to have a pulse duration of 360 ns for a $\pi$ rotation for pulses in the sequence and 750 ns for the detection.
These values were a compromise between leakage---if pulses are too short they have a wider frequency bandwidth---and inhomogeneity---if pulses are too long their duration becomes comparable to $T_2^*$.
Leakage to other transitions outside the qutrit subspace was mitigated by the bandwidth of the excitation pulses and by the probe-head bandwidth, which included only the relevant transition frequencies $f_{01}$ and $f_{12}$. 
The detected in-phase and quadrature components of the echo were analysed with a classical Fourier Transform (FT) and a phase correction.
The experimental value is the amplitude of the FT at the resonance frequency.

% QUTRIT STATE TOMOGRAPHY
\subsection*{Qutrit state tomography}
\noindent
We performed a complete tomography of the final state after the implementation of the QFT through a set of measurable quantities $q$ from which the density matrix can be reconstructed.
Our observable was the magnetization of the molecular spin qudit along the $\hat{z}$ axis, which generated the RF signal detected by the spectrometer.
Thus, our tomography procedure translated each $q$ into the amplitude of this signal: each $q$ value was extracted one at a time by implementing the QFT, storing the specific $q$ as a population difference between two states $\ket{\mu}$ and $\ket{\nu}$ with additional pulses (see Tab. I), and then detecting this difference with a Hahn-echo sequence incoherent with the QFT sequence.
This process was averaged 1000 times for each $q$.
The fact that AWG and spectrometer were mutually incoherent ensured that any echoes generated by the interplay between pulses sent by different instruments were averaged out on signal accumulation.
Thus, only the echo generated by the spectrometer pulses was detected, proportional to the quantity $q$ stored as a population difference.
The set of quantities $q$ required to perform a full qutrit tomography is reported in Table I, along with the pulse sequences that transform each $q$ into a population difference.

In the case of the refocused QFT, the set of $q$ values were extracted at the moment of the last refocused state (before the last $\pi$ pulse, see Fig. \ref{fig3new}c), for both sequences A and B, each one allowing for the detection of different density matrix elements.
To recover the results of the full QFT operation to be compared with the ideal QFT ($U_d$) and the non-refocused QFT sequence, we have therefore included the effect of the last $\pi$ pulse and reversed the swapping between states when reconstructing the matrix elements from the measured $q$ values.
It is worth noting that this swapping between states can be easily reverted during the actual implementation of the sequence, once the QFT is integrated as part of a larger algorithm.
Matrix elements that are measured by both sequence A and B are averaged, mitigating any possible offset error.

% TABLE 1: Tomography
\begin{table}[h!] \label{tab: ncoding_matrix_elements}
\begin{tabular}{l|l|l}
$q$ &Pulses ($\theta$, $\phi$, $f_{\mu\nu}$) &Detection ($f_{\mu\nu}$) \\ 
\hline \hline
$\rho_{00} - \rho_{11}$ & - &$f_{01}$ \\ \hline
$\rho_{11} - \rho_{22}$ & - &$f_{12}$\\ \hline
$2\, \rm{Re} (\rho_{01})$ & ($\pi/2$, $-\pi/2$, $f_{01}$) &$f_{01}$\\ \hline
$2\, \rm{Im} (\rho_{01})$ & ($\pi/2$, $\pi$, $f_{01}$) &$f_{01}$\\ \hline
$2\, \rm{Re} (\rho_{12})$ & ($\pi/2$, $-\pi/2$, $f_{12}$) &$f_{12}$\\ \hline
$2\, \rm{Im} (\rho_{12})$ & ($\pi/2$, $\pi$, $f_{12}$) &$f_{12}$\\ \hline
$2\, \rm{Re} (\rho_{02})$ & ($\pi$, $-\pi/2$, $f_{12}$), &$f_{01}$\\
& then ($\pi$, $-\pi/2$, $f_{01}$), & \\
& then ($\pi/2$, $\pi/2$, $f_{01}$) & \\ \hline
$2\, \rm{Im} (\rho_{02})$ & ($\pi$, $-\pi/2$, $f_{12}$), &$f_{01}$\\
& then ($\pi$, $\pi$, $f_{01}$), & \\
& then ($\pi/2$, $0$, $f_{01}$) & \\ \hline
\end{tabular}
\caption{
    The eight quantities $q$ required for a full tomography of the $3\times3$ density matrix describing the qutrit state, with the pulses needed to transform each $q$ into a population difference and the frequency (transition) in which this difference is detected with a Hahn-echo sequence.
    Notation for pulses is ($\theta$, $\phi$, $f_{\mu\nu}$), where $\theta$ is the tilt angle, $\phi$ the phase of the carrier signal and $f_{\mu\nu}$ the carrier frequency.
}
\end{table}

% PSEUDO-PURE STATE GENERATION
\subsection*{Pseudo-pure state generation}
\noindent
Before each QFT implementation, we generated a pseudo-pure state by sending a $\pi/2$-pulse with frequency $f_{12}$, to equipopulate states $\ket{1}$ and $\ket{2}$, and by letting the coherence between these states decay for a time much longer than $T_2$.
The resulting density matrix after this procedure in the $\left\{ \ket{0},\ket{1},\ket{2} \right\}$ subspace has therefore the form 
\begin{equation*}
\rho'_{0-2} = \, \alpha_0 \ket{0} \bra{0} 
\, + \, \dfrac{p_1+p_2}{2} \, \mathbb{1} \, ,
\end{equation*}
with $\alpha_0 = p_0-(p_1+p_2)/2$ and $p_{\eta}$ the initial Boltzmann population of the energy states.
Except for a normalization factor ($\alpha_0$), this state is equivalent to the {\it pure} density matrix $\rho_{0-2} = \ket{0} \bra{0}$.
Indeed, the part of $\rho'_{0-2}$ proportional to the identity in the considered subspace ($\mathbb{1}$) does not produce any signal in our experiment.
It is also worth noting that we can work with a pseudo-pure state restricted to a 3-state subspace of the full 12-state qudit thanks to the symmetry properties of the static Hamiltonian of Eq.(\ref{eq:spin_Hamiltonian}).
When the static magnetic field is applied along the $\hat{x}$ axis, Eq.(\ref{eq:spin_Hamiltonian}) is invariant under a $\pi$ rotation around this same axis.
Conversely, our observable is the magnetization of the molecular spin qudit along $\hat{z}$, which has then zeroes all along the diagonal if written in the eigenbasis of the static Hamiltonian.
Thus, only coherences generated by our pulses on the two addressable transitions can be detected: outside the 3-state subspace the contribution is zero as the trace of the product of a diagonal matrix (describing the thermal populations of the other 9 states) and a matrix with all zeroes in the diagonal (the observable) is zero.\\

% NORMALIZATION OF TOMOGRAPHY EXPERIMENTS
\subsection*{Normalization of tomography experiments}
\noindent
The measurement of the echo on transition $\ket{\mu} \leftrightarrow \ket{\nu}$ after a Hahn-echo sequence with frequency $f_{\mu\nu}$, starting from thermal equilibrium, yields an echo intensity $I^{(\rm{eq})}_{\mu\nu} = C_{\mu\nu} \left[ p^{(\rm{eq})}_{\mu} - p^{(\rm{eq})}_{\nu} \right]$.
Here $C_{\mu\nu}$ is a constant proportional to the matrix element of the magnetization along $\hat{z}$ between the two states, and $p^{(\rm{eq})}_{\mu}$ is the population of state $\ket{\mu}$ in thermal equilibrium.
When we generate the pseudo-pure state, the same transition produces an echo intensity $I^{(\rm{pur})}_{\mu\nu} = C_{\mu\nu} \left[ p^{(\rm{pur})}_{\mu} - p^{(\rm{pur})}_{\nu} \right]$, now with $I^{(\rm{pur})}_{12} \propto p^{(\rm{pur})}_1 - p^{(\rm{pur})}_2 \simeq 0$.
After manipulating the pseudo-pure state and converting a quantity $q$ into a population difference, the value of $q$ is obtained from the corresponding echo intensity as
\begin{equation*}
    q = p_{\mu} - p_{\nu} = \dfrac{1}{\alpha} \dfrac{ I_{\mu\nu} }{ I^{(\rm{eq})}_{\mu\nu} }
    \left[ p^{(\rm{eq})}_{\mu} - p^{(\rm{eq})}_{\nu} \right] \,,
\end{equation*}
which depends on a normalization factor $\alpha$.
This factor, obtained from the measurement of $I^{(\rm{pur})}_{01}$ and $I^{(\rm{pur})}_{12}$, ensures that the trace of the density matrix representing the pseudo-pure state---and also of the density matrices representing the manipulation of this state---equals 1 and that all elements in its diagonal are positive.
For an ideal pseudo-pure state with $I_{12}^{(\rm{pur})} = 0$, the normalization factor is
\begin{equation*}
    \alpha = \dfrac{ I^{(\rm{pur})}_{01} }{ I^{(\rm{eq})}_{01} }
    \left[ p^{(\rm{eq})}_0 - p^{(\rm{eq})}_1 \right] 
    = \left[ p^{(\rm{pur})}_0 - p^{(\rm{pur})}_1 \right]  = \alpha_0 \, .
\end{equation*}
\\
\noindent
A small $I_{12}^{(\rm{pur})}$ leads to a correction in the normalization factor,
\begin{equation*}
    \alpha = \alpha_0
    + \dfrac{1+ 3\,\text{sgn}(\epsilon)}{2} \, |\epsilon|
    \left[ p^{(\rm{eq})}_1 - p^{(\rm{eq})}_2 \right] \,,
\end{equation*}
with $\epsilon = I_{12}^{(\rm{pur})}/I_{12}^{(\rm{eq})}$.
Thus, all measurements are normalized by measuring the echo intensities of the thermal equilibrium and pseudo-pure states for both transitions, plus a temperature measurement ($T = 1.4$ K) to compute the population differences in thermal equilibrium: $p^{(\rm{eq})}_0 - p^{(\rm{eq})}_1 = 7.705 \cdot 10^{-4}$, $p^{(\rm{eq})}_1 - p^{(\rm{eq})}_2 = 8.343 \cdot 10^{-4}$.\\

%%%%%%%%%%%%%%%%%%%%%%%%%%
% ACKNOWLEDGMENTS %%%%%%%%
%
\section*{Acknowledgments}
\noindent 
The Authors acknowledge financial support by the European Union - NextGenerationEU, within Mission 4 Component 2, PNRR MUR project PE0000023-NQSTI - “National Quantum Science and Technology Institute” (NQSTI), CUP
D93C2200094000 (M.R.O., L.B., G.A., R.D.R., E.G., S.C.), and within Mission 4 Component 1 within the PRIN 2022 program - project “CROQUET” - 2022L57S28, CUP D53D23010400006 (L.B., G.A., E.G.).
The work was also funded by the Horizon Europe Programme within the ERC-Synergy project CASTLE (proj. n. 101071533) (L.B., S.Ch., G.A., R.D.R., E.G., S.C.) and by the Novo Nordisk Foundation under grant NNF21OC0070832 in the call “Exploratory Interdisciplinary Synergy Programme 2021” (S.Ch., S.P., E.G., S.C.).

%%%%%%%%%%%%%%%%%%%%%%%%%%
% CONTRIBUTIONS %%%%%%%%%%
%
\section*{Author contributions}
\noindent 
S.C. and E.G. conceived the experiment, performed by M.R.O., L.B., S.Ch. and G.A., after discussion with R.D.R..
The single crystal sample was synthesized by A.M., M.D.R. and S.P..
Data analysis and simulations were made by M.R.O. and L.B..
S.C., E.G., and S.P. supervised the project.
M.R.O., L.B., E.G. and S.C. wrote the manuscript with inputs from all co-authors.

%%%%%%%%%%%%%%%%%%%%%%%%%%
% BIBLIOGRAPHY %%%%%%%%%%%
%
% Create the reference section using BibTeX
\bibliographystyle{naturemag}
\bibliography{biblio}
%
%
%%%%%%%%%%%%%%%%%%%%%%%%%%
% SUPP. INFO MERGING %%%%%
%%%%%%%%%% Merge with supplemental materials %%%%%%%%%%
\pagebreak
\widetext
\begin{center}
\textbf{\large SUPPLEMENTARY INFORMATION \\
Implementation of the Quantum Fourier Transform on a molecular qudit with full refocusing and state tomography}
\end{center}
%%%%%%%%%% Merge with supplemental materials %%%%%%%%%%
%%%%%%%%%% Prefix a "S" to all equations, figures, tables and reset the counter %%%%%%%%%%
\setcounter{equation}{0}
\setcounter{figure}{0}
\setcounter{table}{0}
\setcounter{page}{1}
\makeatletter
\renewcommand{\theequation}{S\arabic{equation}}
\renewcommand{\thefigure}{S\arabic{figure}}
\renewcommand{\bibnumfmt}[1]{[S#1]}
\renewcommand{\citenumfont}[1]{S#1}
%%%%%%%%%% Prefix a "S" to all equations, figures, tables and reset the counter %%%%%%%%%%
%
% Authors
\begin{center}
    Marcos Rubín-Osanz$^{1}$,
    Laura Bersani$^{1}$,
    Simone Chicco$^{1}$,
    Giuseppe Allodi$^{1}$,
    Roberto De Renzi$^{1}$,
    Athanasios Mavromagoulos$^{2}$,
    Michael D. Roy$^{2}$,
    Stergios Piligkos$^{2,*}$,
    Elena Garlatti$^{1,3,*}$,
    Stefano Carretta$^{1,3,*}$
\end{center}
\begin{center}
    $^1$Dipartimento di Scienze Matematiche, Fisiche e Informatiche, Universit\`{a} di Parma\\
    and UdR Parma, INSTM, I-43124 Parma, Italy.\\
    $^2$Department of Chemistry, University of Copenhagen, DK-2100 Copenhagen, Denmark.\\
    $^3$INFN, Sezione di Milano-Bicocca, gruppo collegato di Parma, I-43124 Parma, Italy.
\end{center}
\begin{center}
    $^{*}$Corresponding authors. Email: stefano.carretta@unipr.it, elena.garlatti@unipr.it, piligkos@chem.ku.dk
\end{center}
%
%
%%%%%%%%%%%%%%%%%%%%%%%%%%
% SUPP. NOTE 1 %%%%%%%%%%%
\section{Supplementary Note 1: Relaxation measurements and characterization}
\noindent
Figure S1 shows the result of thermal spin lattice relaxation recovery experiments on transitions $\ket{0} \leftrightarrow \ket{1}$ and $\ket{1} \leftrightarrow \ket{2}$.
The initial condition was obtained with a train of pulses equipopulating the two states, followed by an Hahn echo detection ($\frac{\pi}{2} - \pi$ sequence) with a variable $\tau$ delay, during which the system start to recover.
We obtained relaxation times of at least 2 ms for the fastest relaxing spins, two orders of magnitude longer than the duration of our sequences implementing the QFT.
Conversely, the waiting time $t \gg T_2$ to let the coherence between states $\ket{1}$ and $\ket{2}$ decay during the pseudo-pure state generation procedure was long enough to introduce a slight error relative to their equipopulation due to spin-lattice relaxation.
We compensated this error by fine tuning the duration of the equipopulating pulse, ideally a $\pi/2$ pulse without relaxation, until we minimized the population difference between $\ket{1}$ and $\ket{2}$.

For the measurement of the phase memory time $T_2$ (reported in Fig. 1c of the main text), a simple echo sequence (with equal pulses $\frac{2\pi}{3}-\frac{2\pi}{3}$, to maximize the signal while irradiating the same frequency band) was implemented, varying the interval between the two pulses.
Since the shape of the decay is Gaussian, the extracted $T_2$ was defined as the value at which the signal amplitude reached $1/e$ of its maximum value. 

Rabi oscillations (reported in Fig. 1d of the main text) were sampled by sending a variable-$\theta$ pulse, whose duration was finely increased, followed by a refocusing $\pi$ pulse in a spin-echo sequence.

% FIGURE S1
\begin{figure}[h!] 
	\begin{centering}
    \includegraphics[width=0.8\textwidth]{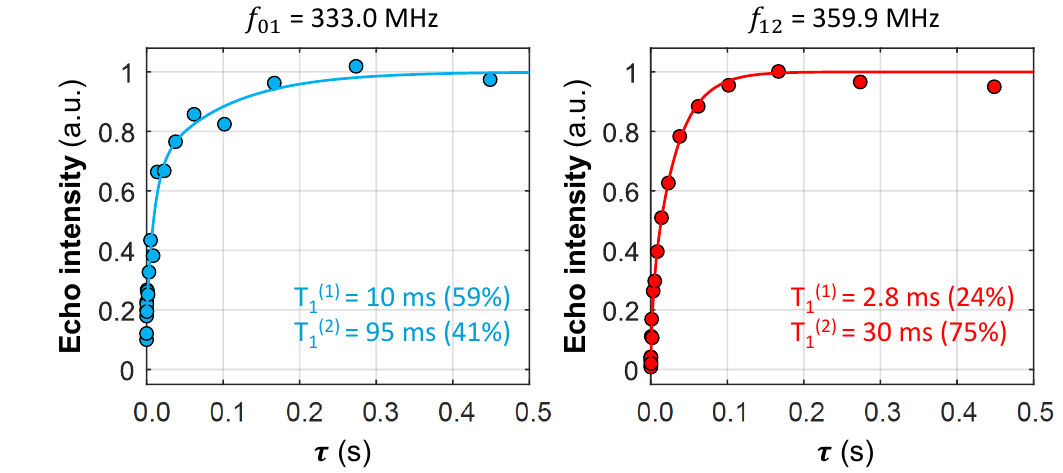}
    	\caption{{\bf Relaxation of the two addressable qutrit transitions.}
            \emph{Left}: relaxation of the $\ket{0} \leftrightarrow \ket{1}$ transition, with frequency $f_{01} =$ 333.0 MHz.
            \emph{Right}: relaxation of the $\ket{1} \leftrightarrow \ket{2}$ transition, with frequency $f_{12} =$ 359.9 MHz.
            In both cases, relaxation is best fit as a bi-exponential decay, with characteristic times $T_1^{(1)}$ (fast) and $T_1^{(2)}$ (slow), which is not unusual for a multilevel system with different possible relaxation pathways.
            The relative weight of both contributions is indicated in $\%$.
        }
	\end{centering}
\end{figure}
\newpage

%%%%%%%%%%%%%%%%%%%%%%%%%%
% SUPP. NOTE 2 %%%%%%%%%%%
\section{Supplementary Note 2: Simulation and origin of inhomogeneous broadening}
\noindent
In order to have a good estimate of the effect of inhomogeneous broadening in the results, we simulated the Hahn-echo sequence for each transition and compared the result with the measured thermal equilibrium state echoes (Fig. S2).
We introduced inhomogeneous broadening as a normal distribution in a parameter of the Hamiltonian---Eq.(1) in the main text---describing the states and energies of the spin system. The resulting magnetization traces for the different spins were averaged according to the weights of this distribution.
Indeed, this generated an echo after a $\pi/2$-$\pi$ sequence of pulses, with the duration of the echo depending on the effect of the chosen parameter on the states involved in the transition.
We found that the origin that better describes the relative duration of the echoes in the two addressed transitions is not a distribution in the parameters of the electronic Zeeman term but a distribution in the hyperfine couplings, with $|\sigma_A/A| = 0.15\%$.
In fact, introducing a distribution in the Zeeman term leads to a significantly shorter echo duration for transition $\ket{1} \leftrightarrow \ket{2}$, as $\Big| \tfrac{df_{12}}{dB} \Big| > \Big| \tfrac{df_{01}}{dB} \Big|$, which we do not observe.

% FIGURE S2
\begin{figure}[h!]
	\begin{centering}
		\includegraphics[width=0.6\textwidth]{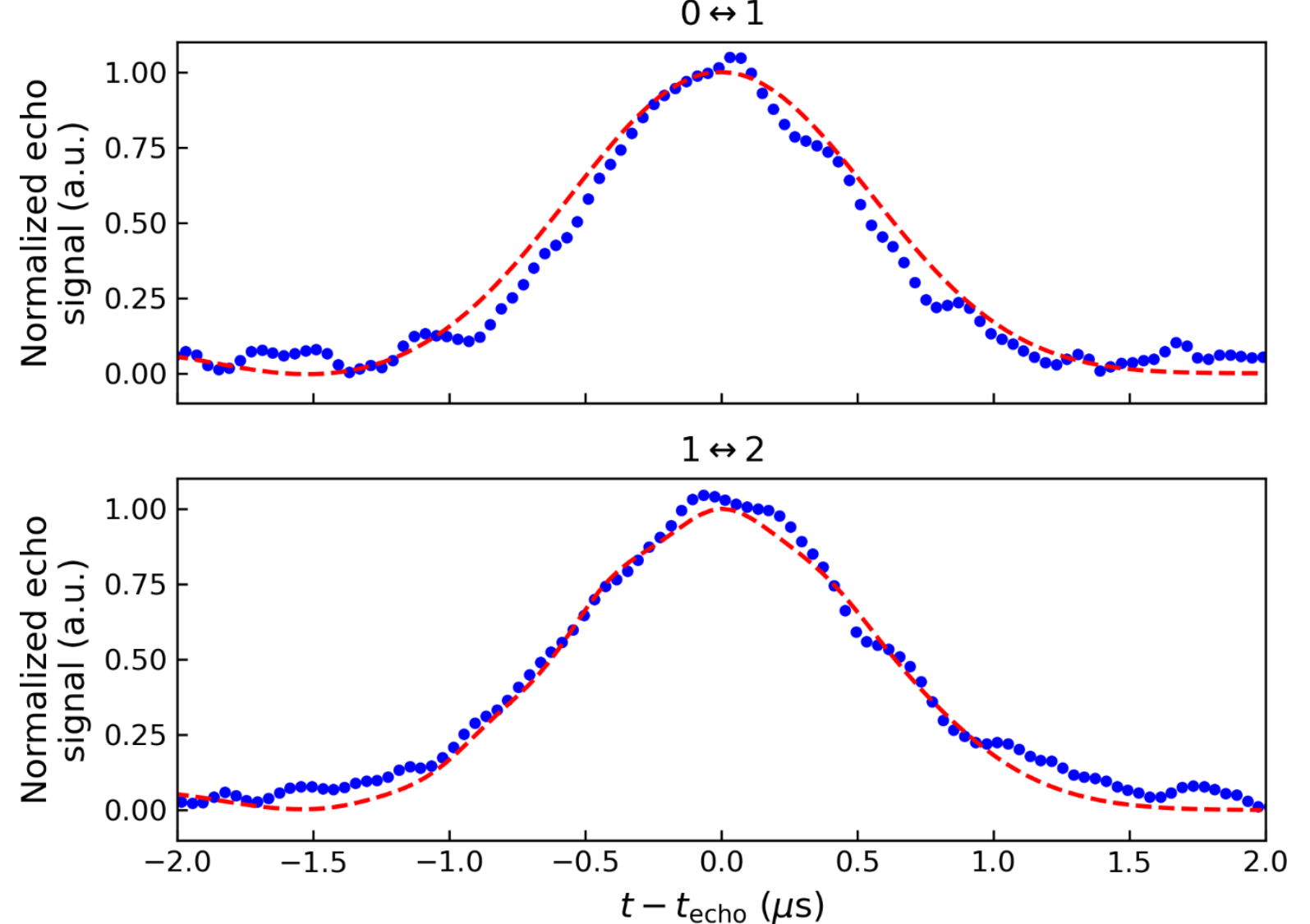}
		\caption{{\bf Thermal echo in the two addressable transitions.}
            Comparison between the measured (blue dots) and simulated (red dashed-lines) echoes.
            \emph{Top}: transition $\ket{0} \leftrightarrow \ket{1}$.
            \emph{Bottom}: transition $\ket{1} \leftrightarrow \ket{2}$.
        }
	\end{centering}
\end{figure}

The effect of non-instantaneous pulses is also included in the simulation.
We found that the relatively long pulses used by the spectrometer in the Hahn-echo detection (750 ns for a $\pi$-pulse, compared to 360 ns with the AWG in the QFT sequence) broaden the measured echo, which leads to an overestimation by eye of the characteristic decay time due to inhomogeneity ($T_2^*$). 
\newpage 

%%%%%%%%%%%%%%%%%%%%%%%%%%
% SUPP. NOTE 3 %%%%%%%%%%%
\section{Supplementary Note 3: Simulation of QFT sequences and tomography experiments}
\noindent
We simulated the non-refocused and refocused QFT pulse sequences implementing the QFT using a distribution with the same strain $|\sigma_A/A|$ determined in the Supplementary Note 2 to describe inhomogeneous broadening.
After each sequence, we also simulated the pulses that transfer the different $q$ to population differences, as well as their incoherent Hahn-echo detection with the spectrometer.
Figure S3 shows the result of these simulations with the three basis states of the qutrit as initial states.
The simulation of the non-refocused sequence yields an attenuation of coherences, markedly on the coherence between $\ket{1}$ and $\ket{2}$ and the two-quanta coherence between $\ket{0}$ and $\ket{2}$, as observed in the experiment.
These results suggests inhomogeneous broadening as the main source of error in the implementation of the QFT with the non-refocused sequence.
Conversely, the simulation of the refocused sequence gives an almost perfect implementation of the QFT (in the absence of other sources of error apart from inhomogeneous broadening).

% FIGURE S3
\begin{figure}[h!]
	\begin{centering}
		\includegraphics[width=1.0\textwidth]{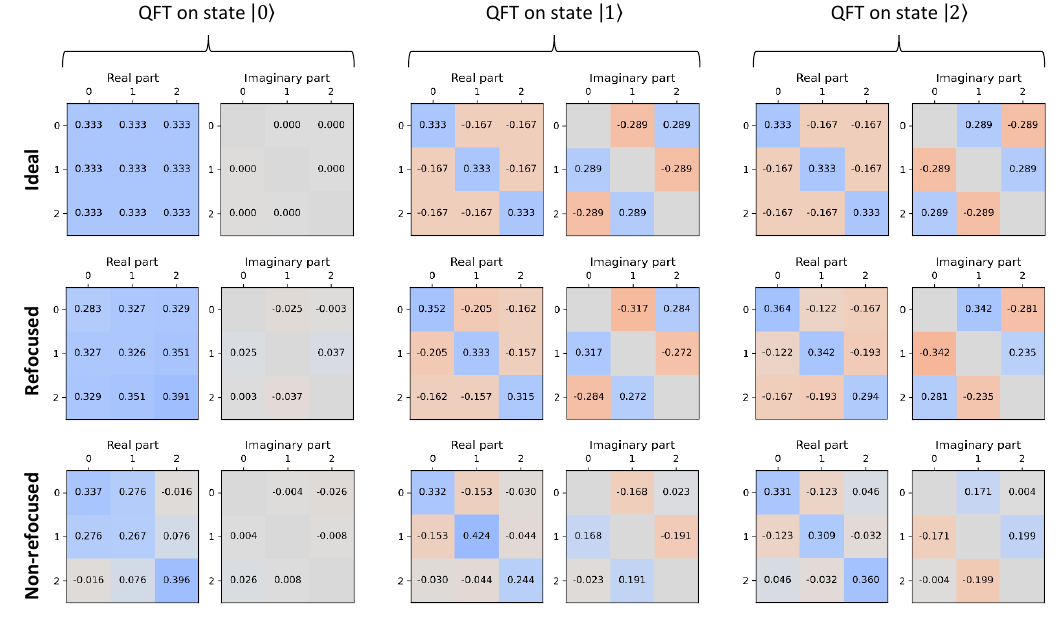}
		\caption{{\bf Comparison between the ideal QFT and the simulation of the refocused and non-refocused implementation on initial states $\ket{1}$ and $\tfrac{1}{\sqrt{2}} \left( \ket{1} - i \ket{2} \right)$.}
            Simulations were carried out with experimental parameters, including only inhomogeneous broadening as a source of error.
            The implementation of the QFT is almost perfect with the refocused sequence, while the non-refocused sequence yields final states with attenuated coherences.
        }
	\end{centering}
\end{figure}
\newpage 

%%%%%%%%%%%%%%%%%%%%%%%%%%
% SUPP. NOTE 4 %%%%%%%%%%%
\section{Supplementary Note 4: Experimental data of the tomography of the initial states}
\noindent
We report in Fig. S4 the tomography of four out of the five initial states used to test our QFT sequences: $\ket{0}$, $\ket{1}$, $\ket{2}$ and the superposition state $\tfrac{1}{\sqrt{2}} \left(\ket{1} - i\ket{2} \right)$.
The tomography for the fifth initial state, $\tfrac{1}{\sqrt{2}} \left(\ket{0} - i\ket{1} \right)$, is shown in Fig. 2b of the main text.
For each initial state, we compare the density matrix of the ideal state ($\rho_{\rm ideal}$) with the density matrix we recover from tomography experiments after the experimental manipulation of the pseudo-pure state ($\rho_{\rm exp}$), obtaining fidelities $\mathcal{F} = \rm{Tr} \left( \sqrt{ \sqrt{\rho_{\rm ideal}} \rho_{\rm exp} \sqrt{\rho_{\rm ideal}}} \right)$ between 0.95 and 0.98 for the initial state generation step.
The error in $\mathcal{F}$ was estimated from the noise in the echo traces following the analysis of Figs. 1 and 2c in the main text.

% FIGURE S4
\begin{figure}[h!]
	\begin{centering}
		\includegraphics[width=1.0\textwidth]{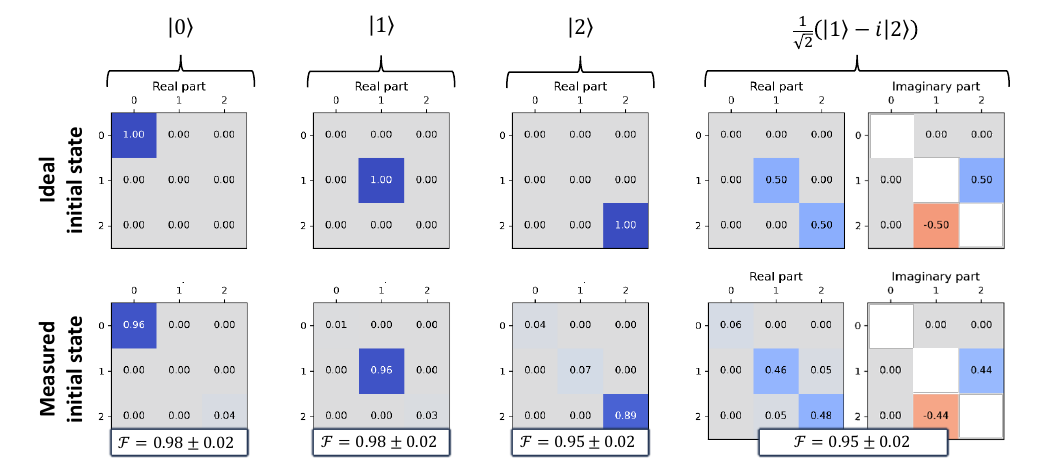}
		\caption{{\bf Tomography of the initial states for the QFT.}
            Initial states were generated from the pseudo-pure state, here shown as the measured initial state $\ket{0}$, with sequences combining $\theta=\pi$ and $\theta=\pi/2$ pulses on the two addressable transitions.
            The tomography of the fifth initial state $\tfrac{1}{\sqrt{2}} \left(\ket{0} - i\ket{1} \right)$ can be found in Fig. 2b of the main text.
        }
	\end{centering}
\end{figure}
\newpage 

%%%%%%%%%%%%%%%%%%%%%%%%%%
% SUPP. NOTE 5 %%%%%%%%%%%
\section{Supplementary Note 5: Experimental data of the tomography of the states after the implementation of the QFT}
\noindent
Here we report those measurements from our full set of tomography data after the implementation of the QFT that were not included in the main text.
Figure S5 compares the non-refocused and refocused QFT implementations for the third basis state, $\ket{1}$.
The fidelity of experimental implementation of the QFT is calculated as $\mathcal{F} = \rm{Tr} \left( \sqrt{ \sqrt{\rho_{\rm ideal}} \rho_{\rm exp} \sqrt{\rho_{\rm ideal}}} \right)$, where $\rho_{\rm ideal}$ is the density matrix obtained by applying $U_d$ to the initial state obtained from a tomography experiment (see Fig. S4 in the Suppplementary Note 4), and $\rho_{\rm exp}$ is the density matrix obtained from the tomography of the state after sending the QFT sequence (refocused or non-refocused).
With the non-refocused sequence we obtain a fidelity $\mathcal{F}=0.90\pm0.01$, which improves to $\mathcal{F} = 0.95\pm0.02$ using the refocused QFT sequence.
Figure S5 also shows the the result of the refocused QFT on the superposition state $\tfrac{1}{\sqrt{2}} \left( \ket{1} - i \ket{2} \right)$, with a fidelity $\mathcal{F}=0.97\pm0.01$.

% FIGURE S5
\begin{figure}[h!]
	\begin{centering}
		\includegraphics[width=0.9\textwidth]{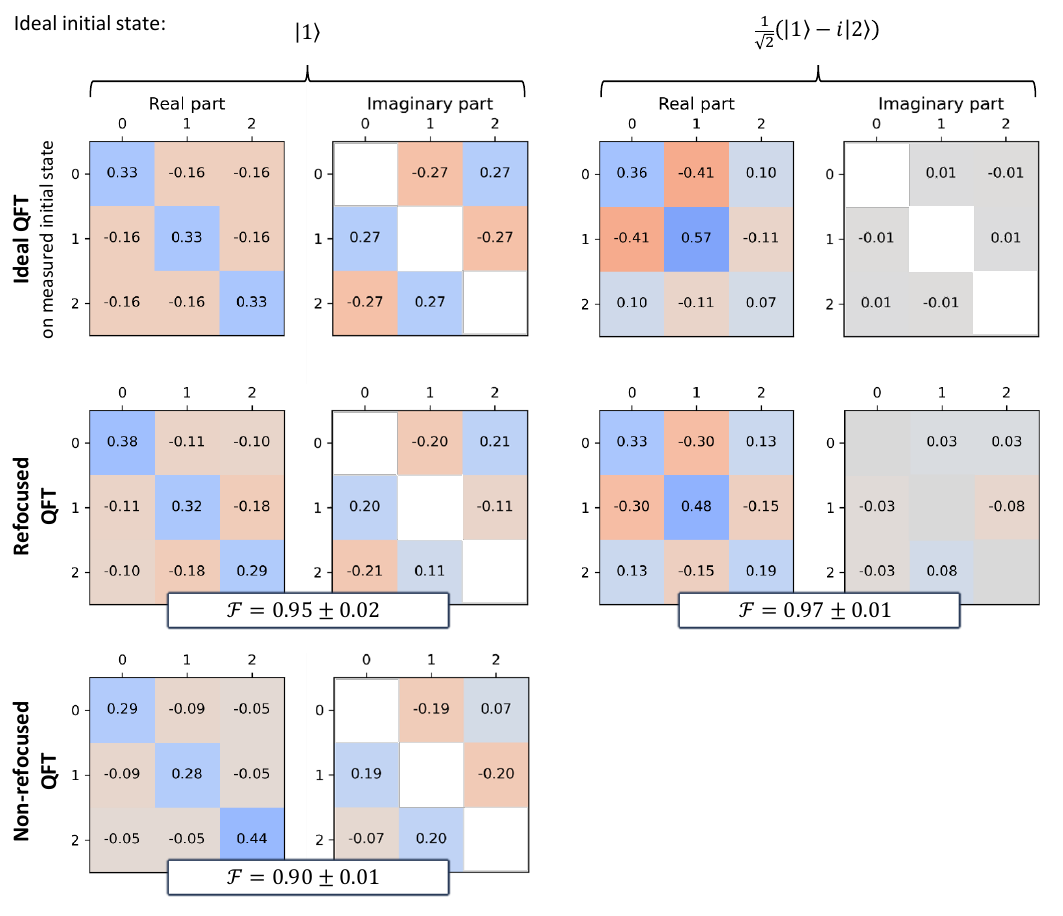}
		\caption{{\bf Comparison between the ideal QFT ($U_d$) on the experimental initial states and its experimental implementation (refocused and non-refocused) on initial states $\ket{1}$ and $\tfrac{1}{\sqrt{2}} \left( \ket{1} - i \ket{2} \right)$.}
            First row was obtained by applying $U_d$ to the tomography of the initial states shown in the bottom row of Fig. S4 in Supplementary Note 4.
        }
	\end{centering}
\end{figure}
\end{document}